\documentclass[apsjnl,amsmath,amssymb,twocolumn,showpacs,superscriptaddress]{revtex4}%
\usepackage{amsmath}
\usepackage{xcolor}
\usepackage{graphicx}
\usepackage{dcolumn}
\usepackage{bm}
\usepackage{float}
\usepackage{epsfig}
\usepackage{graphicx}%
\usepackage{amsmath}%
\usepackage{amsfonts}%
\usepackage{amssymb}
\usepackage{natbib}
\usepackage{soul}
\usepackage{color}

\usepackage{amssymb}
\usepackage{graphicx}
\begin{document}
\title{Clusters of cavity solitons bounded by conical radiation}

\author{Carles Mili\'{a}n}
\email{carles.milian@icfo.eu}
\affiliation{ICFO--Institut de Ciencies Fotoniques, The Barcelona Institute of Science and Technology, 08860 Castelldefels (Barcelona), Spain}
\author{Yaroslav V. Kartashov}
\affiliation{ICFO--Institut de Ciencies Fotoniques, The Barcelona Institute of Science and Technology, 08860 Castelldefels (Barcelona), Spain}
\affiliation{Institute of Spectroscopy, Russian Academy of Sciences, Troitsk, Moscow, 108840, Russia}
\author{Dmitry V. Skryabin}
\affiliation{Department of Physics, University of Bath, Bath BA2 7AY, UK}
\affiliation{ITMO University, St. Petersburg 197101, Russia}
\author{Lluis Torner}
\affiliation{ICFO--Institut de Ciencies Fotoniques, The Barcelona Institute of Science and Technology, 08860 Castelldefels (Barcelona), Spain}
\affiliation{Universitat Polit\`{e}cnica de Catalunya, 08034 Barcelona, Spain}
%
%
%
%
%
\begin{abstract}
We introduce a new class of self-sustained states, which may exist as single solitons or form multi-soliton clusters, in driven passive cylindrical microresonators. Remarkably, such states are stabilized by the radiation they emit, which strongly breaks spatial symmetry and leads to the appearance of long polychromatic conical tails. The latter induce long-range soliton interactions that make possible the formation of clusters, which can be stable if their spatial arrangement is non-collinear with the soliton rotation direction in the microcavity. The clusters are intrinsically two-dimensional and, also, spatially rich. The mechanism behind the formation of the clusters is explained using soliton clustering theory. Our results bring fundamental understanding of a new class of multidimensional cavity solitons and may lead to the development of monolithic multi-soliton sources.
\end{abstract}
\maketitle
\textit{Introduction.---}
The advent of suitable materials and experimental techniques to create so-called \textit{ frequency combs} in micro-cavity rings \cite{vahala03,kippenberg2011microresonator} triggered an intense research program addressed at exploring the existence of combs that are stable and spectrally broadband. These two features are found together in self-sustained microring cavity solitons (CSs), even in the presence of higher-order linear and nonlinear effects \cite{brasch16,karpov16,vahala17},
affording a continuously-renewed source of fundamentally new physical phenomena. One of the most striking discoveries in this context was the existence of frequency locked CSs containing resonant radiation due to higher-order dispersion, first reported in Ref. \cite{erk} and further analyzed in \cite{lamont,milianOE,parra,cherenkov,parraBS,vladimirov18}. Remarkably, while radiative leaky mechanisms may be  detrimental for solitons in conservative systems \cite{akhm95,skryrmp}, they may play a strong stabilizing role in microring settings \cite{milianOE,parra,mbe2017,d1}. Resonant radiation arises due to the matching of the CS dispersion relation with that of the linear waves on top of the steady state background \cite{karpman,kivshar,milianOE,d1}, and it manifest itself by the appearance of prominent spectral peaks linked to the parent solitons \cite{akhm95,chang11}.

In higher dimensions, CSs are known to exist in the context of monochromatic light \cite{tlidi94,lord96,firth02,gomila07}, but they were found to be stable only in a small region of the parameter space if no extra effects, such as stabilizing potentials \cite{valc13}, are considered. By and large, in most of the parameter space strong instabilities and chaos have been shown to occur \cite{tlidi17}. Therefore, a fascinating and so far unexplored question arises about whether radiative leakage may stabilize multidimensional micro-CSs, taking into account that these are pulses with quite broad spectrum instead of single-color beams.

In this Letter, we show that bright two-dimensional CSs are stabilized in a passive cylindrical microcavity by the action of the third order dispersion (TOD). Radiation tails of these CSs are inherently polychromatic as they represent spatio-temporal conical radiation \cite{kolesik04}. Various types of conical radiations emitted by nonlinear pulses are routinely observed as transient phenomena in free-space filamentation optics \cite{faccio06,couairon06,durand13,fabio_shock,babus,mol}. Importantly, radiation bursts emitted \textit{aperiodically} \cite{mol,babus} can help to arrest collapse \cite{mol}. Here we show that conical emission can in fact lock into a complex and strictly stationary CS.
Stable two-dimensional CSs with conical tails exist in the parameter domain where CSs with no radiation are highly unstable. Conical radiation sets an intricate landscape for CS interaction that enables the formation of complex cavity soliton clusters (CSCs) containing a finite number of CSs. CSCs are stable only for particular spatial arrangements having no lower-dimensional analogues. Remarkably, spatial structuring as well as dynamical formation of CSCs can be rigorously understood in terms of the presented clustering theory. In addition, instabilities of CSs may be beneficial as they trigger spontaneous formation of CSCs.

\begin{figure}
\begin{center}
\includegraphics[width=.49\textwidth]{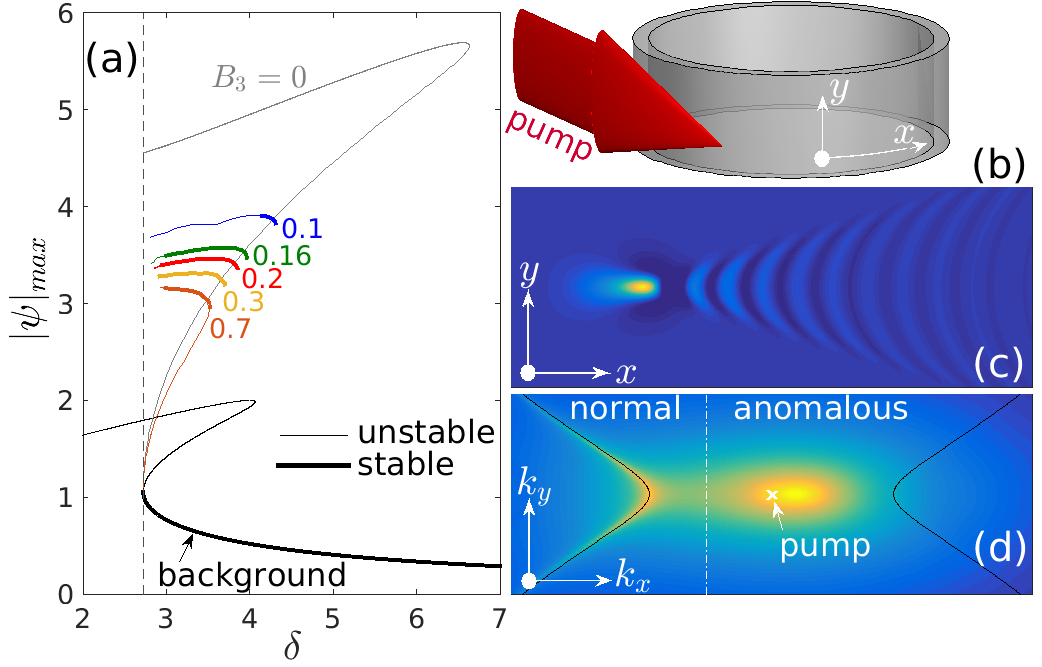}
\caption{(a) CS amplitudes vs $\delta$ and $B_3$ (labels at each branch) around the cavity resonance (black) for $h=2$. Unstable middle branches merging with background are shown only for $B_3=0$ and $B_3=0.7$. Thick (thin) lines denote stable (unstable) states. (b) Sketch of the driven microcavity. An example of stable soliton is shown in spatial (c) and frequency (d) domains for $B_3=0.2$, $\delta=3.3$. In (d), vertical line marks the zero GVD and solid curves mark the conical radiation wavenumbers calculated from Eq. (\ref{eq:con}). Pump is at $k_x$=$k_y$=0. Axes of panels (c), (d) are $x\in[-10,10]$, $y\in[-5,5]$, $k_x\in[-10,10]$, $k_y\in[-15,15]$.\label{f1}}
\end{center}
\end{figure}

\textit{Model.---}
Two-dimensional CSs may exist in cylindrical microcavities, sketched in Fig. \ref{f1}(b), such as micropillars or microtubes \cite{microcylinder, tubular,kipp_tube}. The pump is assumed to excite primarily azimuthal modes with zero group velocity along the vertical coordinate and therefore the corresponding Lugiato-Lefever (LL) model \cite{lugiato87,chembo13} for the intra-cavity field envelope $\psi$ can be written in the form:
\begin{eqnarray}
-i\partial_t\psi=\frac{1}{2}\left(B_2\partial_x^2-2iB_3\partial_x^3+\partial_y^2\right)\psi+(i\gamma-\delta+|\psi|^2)\psi+h
\label{eq1},
\end{eqnarray}
where $t$, $x$, and $y$ are, respectively, the normalized time (in roundtrip units), the periodic azimuthal coordinate, and the translationally invariant vertical coordinate. $B_2\equiv\omega^{(2)}\tau/(2\pi R)^2$, $B_3\equiv\omega^{(3)}\tau/(2\pi R)^3/3!$ are the dispersion coefficients;  $B_2>0$ corresponds to anomalous group velocity dispersion (GVD); $\omega^{(q)}\equiv\partial_\beta^q\omega(\beta)\vert_{\beta_0}$; $\beta$ is the propagation constant; $R$ is the cylinder radius; $\gamma$ accounts for losses; $\delta\equiv[\omega_0-\omega_p]\tau$ is the normalized cavity detuning; $\omega_p$ is the pump frequency; $\omega_0$ is the closest resonance to $\omega_p$; $\tau\equiv2\pi Rn_g/c$ is the roundtrip time for the pump frequency; $n_g$ is the group index at $\omega_{p}$; $Q=\omega_p\tau/\gamma$ is the quality factor; $\psi=E(\tau n_\textrm{NL})^{1/2}$ and $h=(\tau^3{n_\textrm{NL}})^{1/2}\omega_p^2\mathcal{S}/\omega_0$, where $E$, ${n_\textrm{NL}}$, $\mathcal{S}$ are the physical field, nonlinear coefficient, and coupled pump strength. Eq. (\ref{eq1}) is invariant under transformations $\{ta,xa^{1/2},ya^{1/2},\psi a^{-1/2},\gamma a^{-1},\delta a^{-1},ha^{-3/2}\}$. We set $a=\gamma$ and rescale $x\rightarrow x B_2^{1/2}$ that recasts Eq.\ref{eq1} with $B_2=\gamma=1$. The term $+\partial_y^2\psi$ excites light with slow vertical motion corresponding to high-order modes of the cylinder cross section near the cut-off frequency where GVD is typically anomalous (hence the sign $+$) \cite{YulinOL,laags1,laags2,suchkov17,oreshnOE17,d2}. Physical width of the CSC's and that of the pump beam are discussed in the supplemental material \cite{SupMat}, and are of the order of $\sim1$ mm.
%

\textit{Two-dimensional CSs.---}
Solitons with $B_3=0$ are well-known in the context of single-color beams and are stable only in a  narrow parameter region outside the bistability range \cite{lord96,firth02,gomila07}. In cylindrical microcavities, such two-dimensional solitons are found to be strongly stabilized by the action of TOD, $B_3\ne 0$ [Fig. \ref{f1}(a)]. This finding cannot be anticipated by analogy with the one-dimensional case, as the strength and parameter coverage of the soliton instability increase rapidly in the higher dimensions \cite{mynote}. Therefore, our result constitutes  an important step towards the realization of frequency combs in multi-dimensional geometries.

A salient feature of the two-dimensional CSs, crucial for this work, comes from the combination of TOD and transverse GVD. As a result, CSs acquire a conical radiation tail, as shown in Figs. \ref{f1}(c), (d). Such tails fall in the normal GVD regime and therefore resemble free space optical X-waves \cite{conti03}, whose extended tails are interpreted in terms of angular phase matching \cite{kolesik04}. In our case, the spectral content of the conical radiation is obtained by requiring phase matching for linear waves $\sim ae^{ik_x x+ik_y y}$ propagating on top of the $\psi_0$ background away from the soliton core:
\begin{equation}
v_xk_{x}-B_3k_{x}^2=\pm[(\delta+k_y^2/2-|\psi_0|^2)^2-|\psi_0|^4]^{1/2},\label{eq:con}
\end{equation}
where $v_x$ is the nonzero $x$-component of CS velocity (that we obtain together with soliton profile) induced by TOD. Equation (\ref{eq:con}) perfectly describes spectral structure of CS tails, as shown in Fig. \ref{f1}(d) and in Figs. \ref{f3}(c,e) for CSCs.

\textit{Theory of CS interaction and dynamical streams.---}
In order to unveil what types of CSCs exist in cylindrical micro-cavities it is crucial to understand how individual CSs form ensembles. To this end, we develop the cluster formation theory, presented below for the case of two interacting CSs (see further details in the supplemental material \cite{SupMat}). A pair of CSs will bind together at locations in the $xy$-plane meeting equilibria against inter-soliton forces. A superposition of two CSs $\psi_{1,2}$ in the reference frame moving with velocity $v_x$ is accurately described by the ansatz $\psi=\psi_{1}[x_1(t),y_1(t)]+\psi_{2}[x_2(t),y_2(t)]+\chi$, where $\chi$ is a small correction arising due to interacting CSs offset by a distance $[\Delta x^2+\Delta y^2]^{1/2}$ ($\Delta x\equiv x_2-x_1$, $\Delta y\equiv y_2-y_1$). $\dot{x}_{1,2}\equiv\partial_tx_{1,2}$ is the soliton velocity measured relative to $v_x$ and $\dot{y}_{1,2}$ is the transverse velocity component acquired due to CS interaction. When $\psi_{1,2}$ are far from instability thresholds, $\chi$ is most naturally regarded as the superposition of two neutral modes excited due to inter-soliton forces \cite{skrOL,skrvla_pre}. Substituting the above ansatz in Eq. (\ref{eq1}) one obtains linearized equation for $\chi(x,y)$:
\begin{eqnarray}
&& \sum_{q=1}^2\left[\dot{x}_q\partial_{x_q}+\dot{y}_q\partial_{y_q}\right]\vec{S_q}=\hat{\mathcal{L}}\vec{\chi}+\vec{\mathcal{K}},\label{eq3}
\end{eqnarray}
where $\vec{\chi}\equiv[\chi^*, \chi]^\dagger$, $\vec{S_q}\equiv[\psi_q^*-\psi_0^*,\psi_q-\psi_0]^\dagger$, and $\vec{\mathcal{K}}$ contains the terms resulting from the soliton-soliton interactions. Projecting the above equation onto the neutral modes of the operator $\hat{\mathcal{L}}^{\dagger}(\vec{S_p})$, $\eta_p^{(1)}$ and $\eta_p^{(2)}$ ($p=1,2$), leads to the algebraic system of four equations for the soliton velocities $\hat{\mathrm{A}}[\dot{x}_1$, $\dot{x}_{2}$, $\dot{y}_{1}$, $\dot{y}_{2}]^T=\bf{b}$. The elements of the $4\times4$ matrix $\hat{\mathrm{A}}$ and vector $\bf{b}$ are, respectively, the projections of $\eta^{(1,2)}_p$ on the neutral modes of $\hat{\mathcal{L}}(\vec{S_q})$ and on the vector $\vec{\mathcal{K}}$. Two-soliton clusters exist for displacements $\{\Delta x,\Delta y\}$ for which $\Delta v_x\equiv\dot{x}_{1}-\dot{x}_{2}=0$ and $\Delta v_y\equiv\dot{y}_{1}-\dot{y}_{2}=0$. In addition, such clusters are stable against inter-soliton forces only if all velocity vectors $\textbf{e}_x{\Delta v_x}+\textbf{e}_y\Delta v_y$, where $\textbf{e}_x,\textbf{e}_y$ are basis vectors, point towards $\{\Delta x,\Delta y\}$ in close proximity of this point.

\begin{figure}
\begin{center}
\includegraphics[width=.49\textwidth]{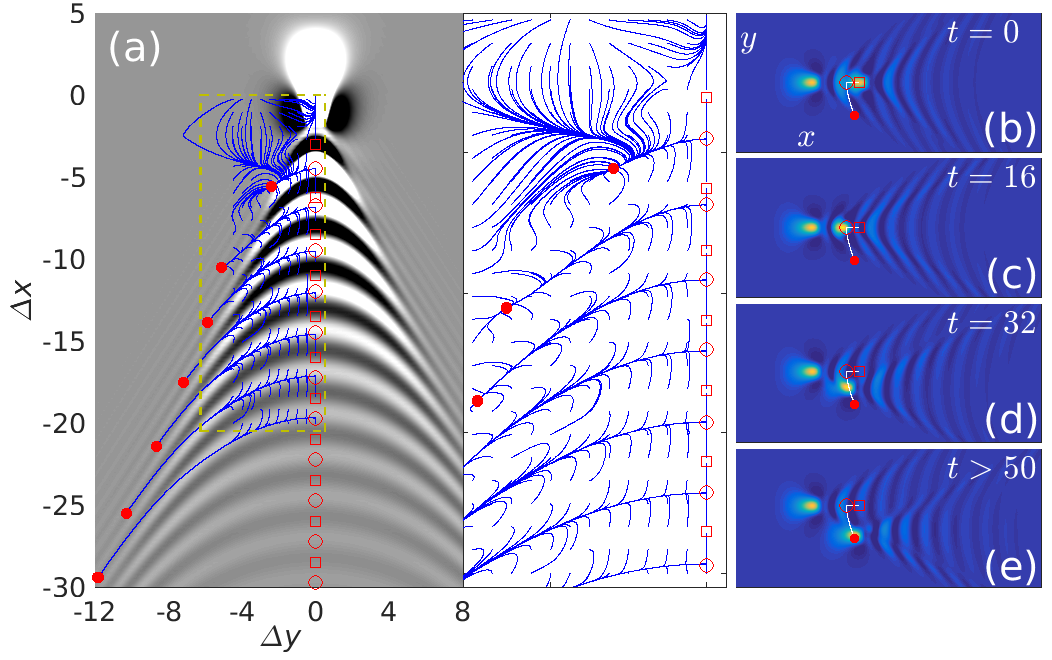}
\caption{(a)
Theoretically predicted displacements corresponding to two-soliton cluster formation, when one solitons is centered at $\Delta x=\Delta y=0$ (gray-scale background): solid circles denote stable equilibria while hollow circles and squares denote unstable equilibria for the locations of the second soliton (see text). Solid curves show predicted streamlines describing relative motion of the two solitons when initial spatial offsets are within the dashed rectangle. Inset in (a) is a zoom of the streamlines in the dashed rectangle. (b-e) Snapshots corresponding to direct numerical propagation of two-soliton cluster corresponding to unstable equilibrium: $\Delta x=-6.25$, $\Delta y=0$. White line shows the associated theoretical streamline. Panel sizes, $x\times y$, are $40\times10$, $B_3=0.7$, $\delta=3.3$.
\label{f2}}
\end{center}
\end{figure}

\begin{figure}
\begin{center}
\includegraphics[width=.49\textwidth]{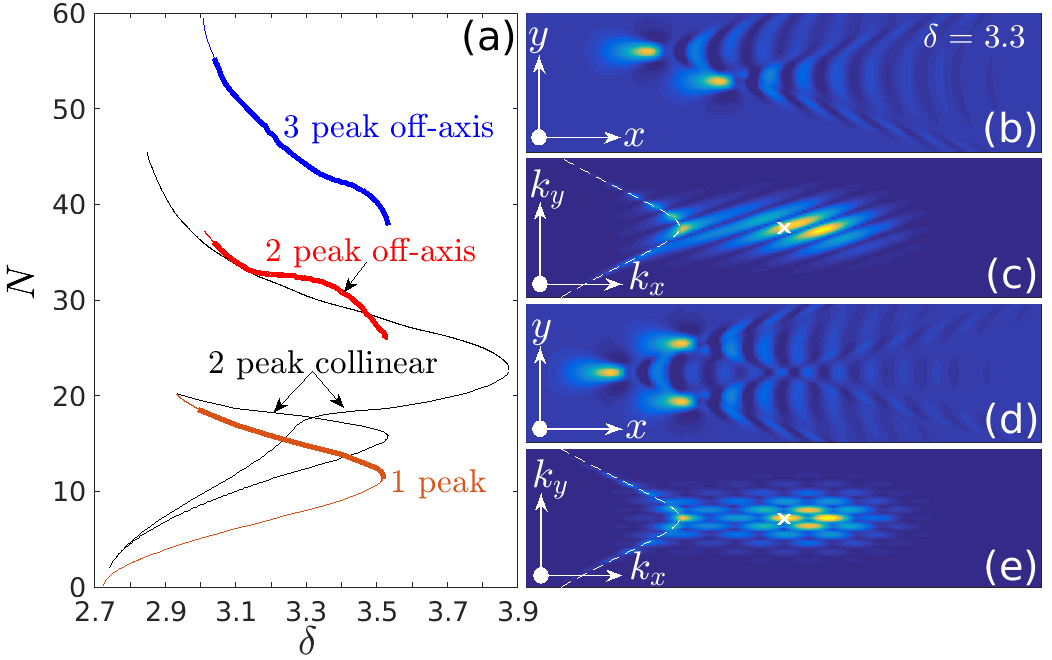}
\caption{(a) Norm $vs$ detuning for single CSs and selected clusters of two and three solitons at $B_3=0.7$. Thick (thin) curves denote stable (unstable) branches. (b-e) Profiles in spatial (b,d) and frequency (c,e) domains of stable clusters with two (b,c) and three (d,e) solitons at $\delta=3.3$. Dashed curves in (c,e) mark the calculated resonant wavenumbers from Eq. (\ref{eq:con}). Axes are: (b,d) $x\in[-20,20]$, $y\in[-5,5]$, and (c,e) $k_x\in[-6,6]$, $k_y\in[-12,12]$.\label{f3}}
\end{center}
\end{figure}

Figure \ref{f2}(a) shows predictions of the above theory for two interacting identical single-peak CSs with $B_3=0.7$ and $\delta=3.3$. The reference soliton is plotted on the background. Displacements at which clusters are found are marked by red dots (stable locations) and hollow circles and squares (unstable locations). We show such displacements only for $\Delta y\leq0$, since the picture is symmetric in $\Delta y$. A first result of this analysis is that all collinear states ($\Delta y=0$) are unstable against inter-soliton interactions, thus our system does not support \textit{stable} analogs of one-dimensional clusters \cite{parraBS,vladimirov18,erk17}. 
Hollow squares correspond to states that are transversely stable but longitudinally unstable: vice versa for hollow circles. Physically, it seems natural that collinear clusters are unstable because a soliton exposed to the radiation of another one interacts nonlinearly with waves having longitudinal and transverse spread of wavevectors. Therefore, any imbalance in frequency mixing processes will favor longitudinal or transverse displacements. In addition to stable locations (red dots), our theory also provides dynamical insight. The relative motion of two interacting solitons can be readily predicted by the streamlines of the vector field $\textbf{e}_x{\Delta v_x}+\textbf{e}_y\Delta v_y$. Figure \ref{f2}(a) shows 250 such streamlines for soliton offsets within the dashed rectangle, which all tend to stable locations. 

Stability and dynamical predictions of interacting solitons have been checked extensively via propagation simulations and found excellent agreement. An example is shown in Figs. \ref{f2}(b-e). In Fig. \ref{f2}(b) we used as an input at $t=0$ the exact (computed numerically) collinear two-soliton cluster that was predicted to be unstable against $\Delta x$ displacements. Because of instability, two peaks approach each other until the state is reached that is unstable only against $\Delta y$ displacements, Fig. \ref{f2}(c). Further propagation leads to the displacement [Fig. \ref{f2}(d)] of the rightmost soliton towards the theoretically predicted stable location [Fig. \ref{f2}(e)], where it remains from $t=50$ to huge times $t>10^4$. Remarkably, the soliton path on the $xy$-plane practically coincides with the theoretical streamline (white solid path). Details on all numerical methods are provided in the supplemental material \cite{SupMat}.

\textit{Clusters of multiple CSs.---}
In the light of the above results, it is natural to expect very rich families of CSCs in cylindrical micro-cavities. Figure \ref{f3}(a) shows selected examples of such families for single-peak CSs and CSCs consisting of two and three CSs. In order to clearly distinguish all families we plot the norm, $N\equiv\iint|\psi(x,y)-\psi_0|\mathrm{d}x\mathrm{d}y$ $vs$ detuning.
Two-CS collinear ($\Delta y=0$) clusters are seen to bifurcate either from the single peak solitons or directly from the cavity background. In Fig. \ref{f3}(a), the latter CSCs have the two CSs with equal amplitudes and thus correspond to the unstable CSCs in Fig.\ref{f2}.

Branches corresponding to CSCs with non-collinear arrangements are shown in Fig. \ref{f3}(a) for the cases of two (solid red) and three (solid blue) solitons. Profiles in spatial and frequency domains are shown for $\delta=3.3$ in Figs. \ref{f3}(b,d) and Figs. \ref{f3}(c,e), respectively. The two-soliton cluster in Fig. \ref{f3}(b) corresponds to the equilibrium point $\Delta x=-5.56$ and $\Delta y=-2.40$ in Fig. \ref{f2}(a). Note that locations $\Delta y=\pm2.40$ are equally favorable for cluster formation. Populating both of them results in the three-soliton cluster in Fig. \ref{f3}(d), therefore its structure is also remarkably well predicted by the theory. Non-collinear two-soliton clusters (including their radiative tails) are clearly asymmetric along the $y$ and $k_y$ axes. Hence transverse recoil effect is possible and leads to the displacement of the spectral maximum associated with soliton into the point with nonzero $k_y$. Due to this, all transversally asymmetric clusters, like the one in Fig. \ref{f2}(e), acquire small transverse velocities $v_y$ ($|v_y/v_x|\lesssim10^{-2}$, see supplemental material \cite{SupMat}) that transform circular orbits into helices, leading to transport of light along the cylinder's axis. On the contrary, transversally symmetric clusters, like those in Figs. \ref{f2}(b,c) and Fig. \ref{f3}(d), have only nonzero longitudinal velocity, $v_x$, keeping circular orbits.

\begin{figure}
\begin{center}
\includegraphics[width=.49\textwidth]{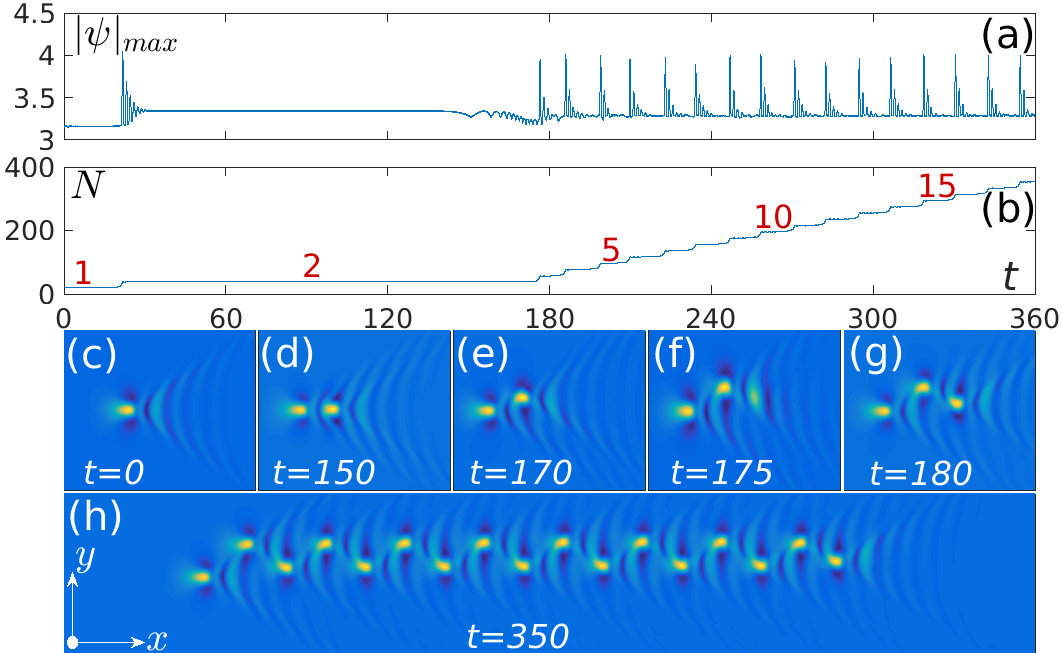}
\caption{Temporal evolutions of (a) peak amplitude and (b) norm of the unstable CS at $B_3=0.7$, $\delta=2.95$. Numbers in (b) indicate number of solitons contained in the pattern. (c-h) Profiles of the intra-cavity field for selected times (see labels) illustrating the zig-zag cluster formation. Panel sizes are ($x\times y$): (c-g) $30\times12$, and (h) $150\times12$.\label{f4}}
\end{center}
\end{figure}

\textit{Instabilities and cluster expansion.---}
Dynamics of the unstable two-dimensional CSs and CSCs is very rich. Figure \ref{f4} illustrates a cascaded process triggered by a single unstable CS at $\delta=2.95$, shown in Fig. \ref{f4}(c). Instability in this case first leads to the spontaneous formation of the two-peak collinear state in Fig. \ref{f4}(d). Such collinear state is unstable, as predicted, and reshapes into an off-axis pair [c.f. Fig. \ref{f4}(e)]. The rightmost intense part of the radiation front eventually gives birth to a third CS that drifts back towards $y=0$. Because the new CS is born with a small transverse drift with respect to its parent, the drift will necessarily be downside until the predicted equilibrium location is reached, in agreement with Fig. \ref{f2}(a). This cascaded process results in the formation of a zig-zag-shaped cluster [Fig. \ref{f4}(h)]. The appearance of new CSs leads to temporal spikes in peak amplitude and ladder steps in norm, as apparent from Figs. \ref{f4}(a) and \ref{f4}(b), respectively. Generation of new CSs is a stimulated process typical when background has inhomogeneities (see, e.g., Refs.\cite{parra14drift, milianPRA}). This cascaded process is arrested when the pattern extends all over the cavity lengths or when the detuning is slightly increased and moved into the stability domain for single CSs. While in the former case the 
pattern becomes chaotic, in the latter case the zig-zag cluster breaks into off-axis pairs and single peak CSs (not shown).

\begin{figure}
\begin{center}
\includegraphics[width=.49\textwidth]{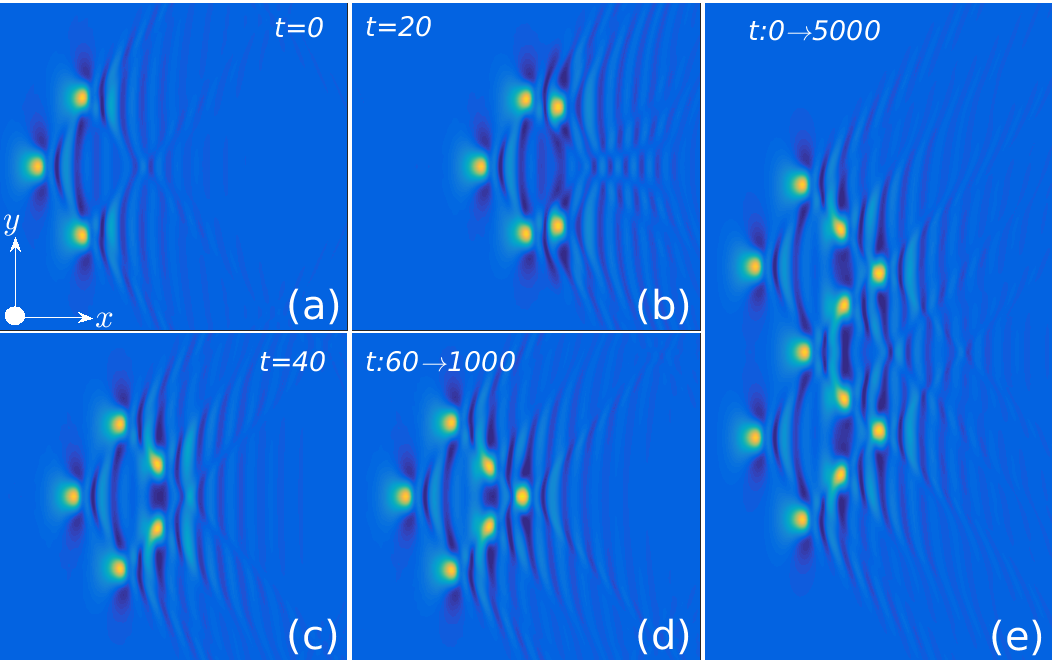}
\caption{(a-d) Formation of a stable 6-peak CSC from an unstable 3-peak one: $B_3=0.7$, $\delta=3.02$. (e) Stable 11-peak cluster built from the CSC in (d). Labels indicate temporal instants (a-c) or intervals (d,e) at which profiles are observed. Panel sizes ($x\times y$): (a-d) $50\times12$, (e) $50\times24$.   \label{f5}}
\end{center}
\end{figure}

Instabilities may result in spontaneous formation of larger stable CSCs. This is a remarkable dynamical feature of this system, as instabilities, easily triggered via cavity detuning, become beneficial for exciting complex states without the need to construct them from individual CSs placed in predetermined locations. Figures \ref{f5}(a-d) show an example of this situation, stimulated by an unstable three-peak cluster at $t=0$ [c.f., Fig. \ref{f5}(a)]. Similarly to dynamics in Fig. \ref{f4}, the intense radiation peaks stimulate the formation of two new solitons (Fig. \ref{f5}(b)) that shift towards the center, Fig. \ref{f5}(c), as dictated by the radiation tails they are exposed to. When they approach each other, as predicted in Fig. \ref{f2}(a) for $\Delta x=0$, they start repelling one another and settle into a stable equilibrium. Moreover, their radiation fronts strongly interfere and a sixth soliton appears at the front, locking the ensemble together to form a stable cluster, represented in Fig. \ref{f5}(d). Interestingly, this complex cluster can be used to form larger stable CSCs, like the eleven peak CSC in Fig. \ref{f5}(e).
%

The existence of stable finite-size CSCs formed by optical pulses with extended radiative tails, c.f. Figs. \ref{f3}(b,d) and Figs. \ref{f5}(d,e), having no one-dimensional analogues, is the central result of this Letter. Until the date, stable patterns of Eq. (\ref{eq1}) where only known with $B_3=0$ and in the form of infinitely extended hexagonal arrangements \cite{firth92,tlidi96}. Our results motivate experimental investigation of suggested structures, which could find applications in multi-channel soliton sources, that do not require structuring of dielectric rods into stacks of microrings.

\textit{Conclusions.---}
We introduced a new class of stable multi-dimensional CS in monolithic cylindrical micro-cavities exhibiting pronounced and polychromatic conical radiation tails. We showed that modulations induced by these tails strongly break CS symmetry and draw a complex effective potential ruling interaction of CSs, which can be understood with the presented soliton clustering theory. Equilibrium points were found to exist at \textit{a priori} counterintuitive spatial locations, leading to intrinsically two-dimensional and highly non-trivial stable clusters with no lower-dimensional analogues. Our results are physically rich and bring fundamental insights into the physics of cavity solitons.

This work was supported by MINECO through the Juan de la Cierva incorporaci\'{o}n program, Severo Ochoa SEV-2015-0522, and grant FIS2015-71559-P; Generalitat de Catalunya, CERCA; Fundaci\'{o} Cellex, Fundaci\'{o} Mir-Puig; The Leverhulme Trust (RPG-2015-456); H2020 (691011, Soliring); RFBR (17-02-00081).


\clearpage
\appendix

\renewcommand\thefigure{\thesection.\arabic{figure}}    
\setcounter{figure}{0}   

\section*{Supplemental Material: Clusters of cavity solitons bounded by conical radiation}

The model under theoretical and numerical analysis is Eq. (1) in the main article:
\begin{eqnarray}
-i\partial_t\psi=\frac{1}{2}\left(B_2\partial_x^2-2iB_3\partial_x^3+\partial_y^2\right)\psi+(i\gamma-\delta+|\psi|^2)\psi+h
\label{eq1}.
\end{eqnarray}

\section{NUMERICAL METHODS}
Three different numerical techniques are used to study propagation in time, stationary solutions, and stability of the latter.
\subsection{Propagation dynamics}
Propagations of the intra-cavity field envelope were carried out by implementing the standard 4$th$ order Runge-Kutta scheme. The spatial derivatives in $x$, $y$ where evaluated in the spectral domain $k_x$ ,$k_y$ by making use of Fourier transforms. Hence, the propagation problem is regarded as:
\begin{eqnarray}
\nonumber && \partial_t\psi=\frac{i}{2}{\mathcal{F}}_{2d}^{-1}\left(\left[-B_2k_x^2+2B_3k_x^3-k_y^2\right]{\mathcal{F}}_{2d}(\psi)\right)-\\ &&
-(\gamma+i\delta-i|\psi|^2)\psi+ih,
\label{eq2},
\end{eqnarray}
where ${\mathcal{F}}_{2d}$ and ${\mathcal{F}}_{2d}^{-1}$ denote, respectively, the direct and inverse two-dimensional Fourier transforms defined as:
\begin{eqnarray}
&& {\mathcal{F}}_{2d}(\psi)\equiv\int_{-\infty}^{+\infty}\mathrm{d}x\mathrm{d}y\ \psi(x,y)e^{i(k_xx+k_yy)}
\\ &&
{\mathcal{F}}_{2d}^{-1}(\tilde\psi)\equiv\int_{-\infty}^{+\infty}\frac{\mathrm{d}k_x}{2\pi}\frac{\mathrm{d}k_y}{2\pi}\tilde\psi(k_x,k_y)e^{-i(k_xx+k_yy)}.
\end{eqnarray}
Numerically, the two above integrals are evaluated with fast Fourier transform routines.
\subsection{Stationary solutions}
Solutions of Eq. \ref{eq1} are sought in the form $\psi(x,y,t)=\phi(X=x-v_xt,Y=y-v_yt,t)$, $\partial_t\phi(X,Y,t)=0$, which leads to:
\begin{eqnarray}
&&
0=M({\phi},{v_x},{v_y})\equiv-i\hat{L}({v_x},{v_y}){\phi}+
|{\phi}|^2{\phi}+h,\label{eq5}
\\ &&\nonumber
\hat{L}({v_x},{v_y})\equiv
{v_x}\partial_X+{v_y}\partial_Y+\\ &&
+\frac{i}{2}B_2\partial_X^2+B_3\partial_X^3+\frac{i}{2}\partial_Y^2-\gamma-i\delta.\label{eq6}
\end{eqnarray}
The solution ${\phi}$ may be characterized by velocity drifts along the longitudinal and transversal directions, denoted by ${v_x}$ and ${v_y}$, which are computed together with ${\phi}$. We use the modified Newton method to solve Eq.\ref{eq5}. Considering that an accurate guess, $\{\bar{\phi},\bar{v_x},\bar{v_y}\}$, of the exact solution $\{\phi,{v_x},{v_y}\}=\{\bar{\phi},\bar{v_x},\bar{v_y}\}+\{\delta\phi,\delta v_x,\delta v_y\}$ is known, the linearized equations in $\delta\phi,\delta v_x,\delta v_y$ are:
\begin{eqnarray}
&& \nonumber
0= M(\bar{\phi}+\delta\phi,\bar{v_x}+\delta v_x,\bar{v_y}+\delta v_y)= M(\bar{\phi},\bar{v_x},\bar{v_y})+\\ && \nonumber
-i\hat{L}(\bar{v_x},\bar{v_y})\delta\phi
+2|\bar{\phi}|^2\delta\phi+\bar{\phi}^2\delta\phi^{*}-i\delta v_x\partial_X\bar{\phi}-i\delta v_y\partial_Y\bar{\phi},\\ &&
\label{eq7}
\end{eqnarray}
which can be cast in matrix form:
\begin{eqnarray}
\nonumber &&
0= \left[\begin{array}{cc}
M' \\
M''
\end{array}\right]+\hat{J}
\left[\begin{array}{cc}
\delta\phi' \\
\delta\phi''
\end{array}\right]
+\delta v_x\partial_X\left[\begin{array}{cc}
\bar{\phi}'' \\
-\bar{\phi}'
\end{array}\right]
+\delta v_y\partial_Y\left[\begin{array}{cc}
\bar{\phi}'' \\
-\bar{\phi}'
\end{array}\right]\\ &&
\hat{J}\equiv
\left[\begin{array}{cc}
\hat{L}''+2|\bar{\phi}|^2+(\bar{\phi}^2)' &\hat{L}'+(\bar{\phi}^2)'' \\
-\hat{L}'
+(\bar{\phi}^2)'' & \hat{L}''+2|\bar{\phi}|^2-(\bar{\phi}^2)'\end{array}\right]\label{eq8},
\end{eqnarray}
where $f'$ and $f''$ denote, respectively, the real and imaginary parts of $f$, and $\hat{J}$ is the Jacobian matrix. At each iteration, $\delta\phi$, $\delta v_x$, $\delta v_y$ are computed and added to the input guess $\bar{\phi}$, $\bar{v_x}$, $\bar{v_y}$, defining input for next iteration. The routine is stopped when $M(\bar{\phi},\bar{v_x},\bar{v_y})=0$ at machine precision. Below, we briefly outline the procedure to compute steady and moving solutions.
\subsubsection{Quiescent solutions}
When the dissipative soliton solution $\bar{\phi}$ is at rest in the frame $x,y$ (i.e., $\gamma>0$, $B_3=0$), $\delta v_x$, $\delta v_y$ are set to zero and the correction to the guess soliton profile is given by:
\begin{eqnarray}
&&
\left[\begin{array}{cc}
\delta\phi' \\
\delta\phi''
\end{array}\right]=-\hat{J}^{-1}\left[\begin{array}{cc}
M' \\
M''
\end{array}\right].
\end{eqnarray}
%
%
\subsubsection{Solitons with longitudinal motion}
When $B_3\neq0$, solitons acquire a velocity $v_x$ in the frame $x,y$. Soliton solutions with well defined symmetry along $y$ have no drift and hence $v_y\equiv0$. In this case, Eq. \ref{eq8} is underdetermined because there are $2N_xN_y$ equations for $2N_xN_y+1$ unknowns ($N_x$, $N_y$ are the number of points along $x$, $y$). By assuming that $\bar{\phi}$ is close to the solution $\phi$, we can fix the real or imaginary parts of the field $\phi$ at the point $(X=X_a,Y=Y_a)$. This amounts to set to zero the component $a\in[1,2N_xN_y]$ of our choice in the correction vector:
\begin{eqnarray}
&&
\left(\left[\begin{array}{cc}
\delta\phi' \\
\delta\phi''
\end{array}\right]\right)_a\equiv 0\label{eq10}.
\end{eqnarray}
Thus the computation of the corrections $\delta\phi$ and $\delta v_x$ at each iterative step is carried out in two steps: First we compute the correction to the velocity,
\begin{eqnarray}
&&
\delta v_x=-\frac{\left(\hat{J}^{-1}\left[\begin{array}{cc}
M' \\
M''
\end{array}\right]\right)_a}
{\left(\hat{J}^{-1}\partial_X\left[\begin{array}{cc}
\phi'' \\
-\phi'
\end{array}\right]\right)_a},\label{eq11}
\end{eqnarray}
and then the correction to the soliton profile, using the $\delta v_x$ from Eq.\ref{eq11}:
\begin{eqnarray}
&&
\left[\begin{array}{cc}
\delta\phi' \\
\delta\phi''
\end{array}\right]=-
\hat{J}^{-1}\left(\left[\begin{array}{cc}
M' \\
M''
\end{array}\right]+\delta v_x\partial_X\left[\begin{array}{cc}
\phi'' \\
-\phi'
\end{array}\right]\right).
\end{eqnarray}
Successful implementation of this method requires that the component $a$ is chosen within the spatial locations where the soliton profile deviates  substantially from the cavity background field.
\subsubsection{Solitons with longitudinal and transverse motion}
When soliton clusters have an asymmetric profiles in $y$, as it is the case for the stable two-soliton bound states, stationary solutions are characterized by two velocities: longitudinal, $v_x$, and transverse, $v_y$. In this case, we proceed analogously to the single velocity problem described in the above section, by noting that Eq.\ref{eq8} contains two unknowns more than equations, i.e., the two velocities $v_x$, $v_y$. Hence, we now choose two components of the correction vector that are assumed to be very small and, in analogy to Eq. \ref{eq10}, write:
\begin{eqnarray}
&&
0\approx-\left(\left[\begin{array}{cc}
\delta\phi' \\
\delta\phi''
\end{array}\right]\right)_{a,b}=\label{eq13}\\ &&
\left(\hat{J}^{-1}\left(\left[\begin{array}{cc}
M' \\
M''
\end{array}\right]+\delta v_x\partial_X\left[\begin{array}{cc}
\phi'' \\
-\phi'
\end{array}\right]+\delta v_y\partial_Y\left[\begin{array}{cc}
\phi'' \\ \nonumber
-\phi'
\end{array}\right]\right)\right)_{a,b}.
\end{eqnarray}
The above system of equations leads to the estimates for the velocities:
\begin{eqnarray}
&&
\delta v_x=\frac{A_2B_1-A_1B_2}{B_2C_1-B_1C_2},\ 
\delta v_y=\frac{A_2C_1-A_1C_2}{C_2B_1-C_1B_2},\label{eq14}
\\  && \nonumber
A_1=\left(\hat{J}^{-1}\left[\begin{array}{cc}
M' \\
M''
\end{array}\right]\right)_a,\ A_2=\left(\hat{J}^{-1}\left[\begin{array}{cc}
M' \\
M''
\end{array}\right]\right)_b,\\ && \nonumber
B_1=
\left(\hat{J}^{-1}\partial_X\left[\begin{array}{cc}
\phi'' \\
-\phi'
\end{array}\right]\right)_a,\ B_2=
\left(\hat{J}^{-1}\partial_X\left[\begin{array}{cc}
\phi'' \\
-\phi'
\end{array}\right]\right)_b,\\ && \nonumber
C_1=
\left(\hat{J}^{-1}\partial_Y\left[\begin{array}{cc}
\phi'' \\
-\phi'
\end{array}\right]\right)_a,\ C_2=
\left(\hat{J}^{-1}\partial_Y\left[\begin{array}{cc}
\phi'' \\
-\phi'
\end{array}\right]\right)_b.
\end{eqnarray}
Once the estimates for the velocities are obtained, the correction to the soliton profile is obtained:
\begin{eqnarray}
&&
\left[\begin{array}{cc}
\delta\phi' \\
\delta\phi''
\end{array}\right]=\label{eq15}\\ &&\nonumber-
\hat{J}^{-1}\left(\left[\begin{array}{cc}
M' \\
M''
\end{array}\right]+\delta v_x\partial_X\left[\begin{array}{cc}
\phi'' \\
-\phi'
\end{array}\right]+\delta v_y\partial_Y\left[\begin{array}{cc}
\phi'' \\
-\phi'
\end{array}\right]\right).
\end{eqnarray}
\begin{figure}
\begin{center}
\includegraphics[width=.49\textwidth]{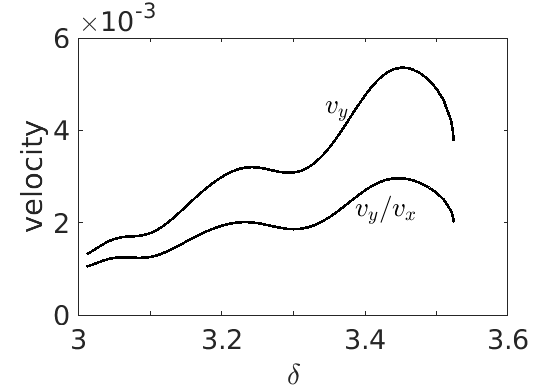}
\caption{Transverse velocity, $v_y$, and transverse to longitudinal velocity ratio, $v_y/v_x$, as a function of cavity detuning for the two-soliton clusters with $B_3=0.7$ [c.f. Fig. 3 in main article].\label{fs1}}
\end{center}
\end{figure}
We note, that conversely to the single velocity problem, the nullity condition in Eq. \ref{eq13} cannot be strictly satisfied for both $a$ and $b$ simultaneously. We found, however, that convergence of this method occurs when $a$ and $b$ are neighboring grid elements. This is important to guarantee that the velocity estimates at each iteration are compatible. In addition, it is important to limit the spatial region of the grid where $a$ and $b$ are chosen as to avoid zero spatial derivatives of $\phi$. The latter makes $B_1$, $B_2$ and/or $C_1$, $C_2$ too close to zero and velocity estimates become inaccurate, causing the divergence of the method.

As an example of the results obtained with the above method, Fig. \ref{fs1} shows the computed velocities $vs$ detuning, $\delta$, for the two-peak off-axis soliton clusters shown in Figs. 3(a-c) in the main article. For this particular example, the magnitude of the transverse velocity varies from $v_x/1000$ to $v_x/300$. Taking into account that $v_x$ itself is already a velocity shift from the group velocity at the pump frequency, $v_y$ is hence rather small. However, in the cylindrical setup considered in this work, any small transverse velocity will cause the cluster to drift way from the center of the pump ($y\approx0$) after a sufficiently long time.
\subsection{Stability of solitons}
While stability is easily tested by propagation simulations, we also performed standard linear stability analysis by substituting in Eq. \ref{eq1} $\psi=\phi(x-v_xt,y-v_yt)+ae^{\lambda t}+b^*e^{\lambda^* t}$, where $\phi$ is the soliton solution, $a$ and $b$ are small constants ($|a|,|b|\ll|\phi|$), and $\lambda$ are the eigenvalues of the associated Jacobian matrix. Instabilities are therefore determined by positive real parts of $\lambda$.
\section{SOLITON CLUSTERS: PERTURBATION THEORY}
%
Equation \ref{eq1} in the rest frame of a solution $\psi$ with arbitrary velocities along $x$ and $y$ reads:
\begin{eqnarray}
&&\partial_t\psi=\hat{L}(v_x,v_y)\psi+K(\psi)+ih\\ &&\nonumber
\hat{L}=v_x\partial_x+v_y\partial_y+\\ &&
+\frac{i}{2}[B_2\partial_x^2+\partial_y^2]+B_3\partial_x^3-\gamma-i\delta\\&&
K(\psi)=i|\psi|^2\psi
\label{eq16}.
\end{eqnarray}
For the analysis below it is useful to remove the background field, $\psi_0$, from the solitons and write $\psi=\psi_0[1+S]$, where $\psi_0$ and $S$ satisfy:
\begin{eqnarray}
&&0=\hat{L}\psi_0+K(\psi_0)+ih\\ &&
\partial_tS=0=\hat{L}S+|\psi_0|^2[K(1+S)-i]
\label{eq17}.
\end{eqnarray}
We now decompose $S=S_{1}[x_1(t),y_1(t)]+S_{2}[x_2(t),y_2(t)]+\chi(x,y)$, where $S_{1,2}$ are two different CSCs and $\chi$ is the small correction accounting for the small reshaping due to the inter-cluster interactions. Subindices 1 and 2 in the coordinates denote that CSs are spatially offset, so $S_1(x_1,y_1)$ and $S_2(x_2,y_2)$ have their peak amplitudes at $x_1=0$, $y_1=0$ and at $x_2=x_1+\Delta x=0$, $y_2=y_1+\Delta y=0$, respectively, so CSs are offset by a distance $[\Delta x^2+\Delta y^2]^{1/2}$. In principle, one can consider $S_1$ and $S_2$ to be arbitrary soliton clusters. However, due to the complexity of the interactions induced by the radiation tails, this theory produces the best results when it is applied to the superposition of two single peak solitons, as shown in Fig. 2 in the main article. In such case, both $S_1$ and $S_2$ are at rest in the same frame, i.e., both are solutions of Eq. 16 with the same $v_x$, $v_y$. Substitution of the above ansatz in Eq. 16 and linearizing in $\chi$ leads to two coupled equations for $\chi(x,y)$ and $\chi(x,y)^*$:
\begin{eqnarray}
&& \sum_{q=1}^2\left[\dot{x}_q\partial_{x_q}+\dot{y}_q\partial_{y_q}\right]\vec{S_q}=\hat{\mathcal{L}}\vec{\chi}+\vec{\mathcal{K}},\label{eq21}\\ &&
\vec{S}_{q}=\left[\begin{array}{c} S_q \\ S_q^*\end{array}\right],\ 
\vec{\chi}=\left[\begin{array}{c} \chi \\ \chi^*
\end{array}\right],\\ &&
\hat{\mathcal{L}}(S_1,S_2)\equiv\\ &&\nonumber\left[\begin{array}{cc}
\hat{L}+2i|\psi_0|^2|1+S_1+S_2|^2 & i|\psi_0|^2[1+S_1+S_2]^2 \\
-i|\psi_0|^2[1+S_1^*+S_2^*]^2 & \hat{L}^*-2i|\psi_0|^2|1+S_1+S_2|^2
\end{array}\right]
\\ &&
\overrightarrow{\mathcal{K}}\equiv\\&& \nonumber|\psi_0|^2
\left[\begin{array}{c} K(1+S_1+S_2)-K(1+S_1)-K(1+S_2)+i \\ K^*(1+S_1+S_2)-K^*(1+S_1)-K^*(1+S_2)-i
\end{array}\right].
\end{eqnarray}
%
\begin{figure*}
\begin{center}
\includegraphics[width=.99\textwidth]{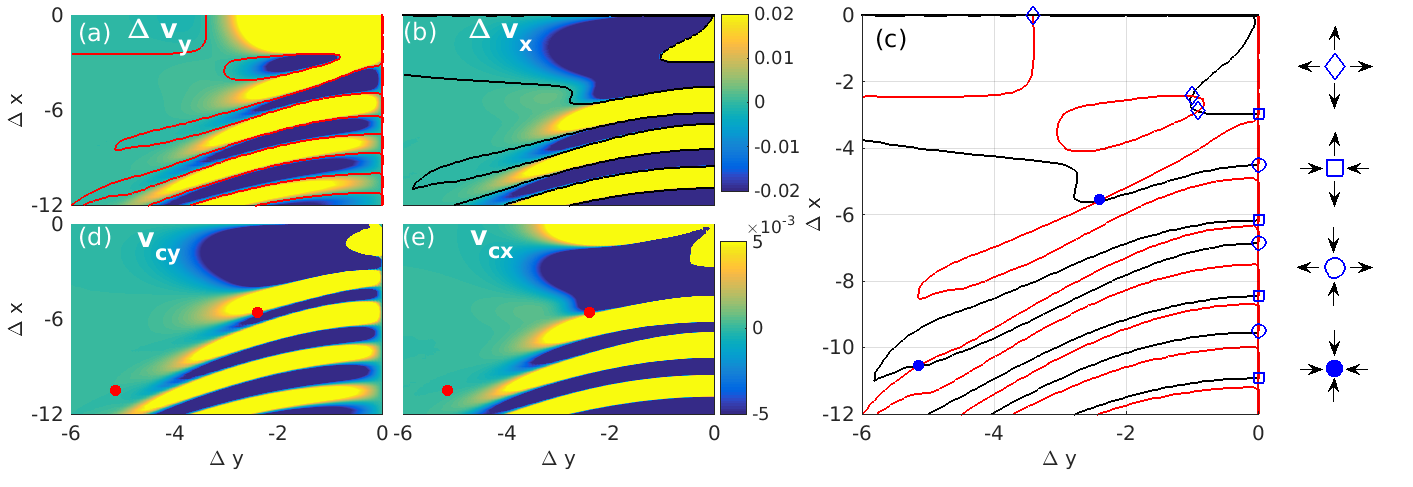}
\caption{Relative soliton velocity along $y$ (a) and $x$ (b) as a function of their separations. Contours mark zero relative speeds. (c) Superpositions of contours in (a) and (b). Intersections represent two-soliton cluster solutions with different stability properties (see text), sketched on the right panel. (d,e) show the fields $v_{cx}=[\dot{x}_{1}+\dot{x}_{2}]/2$, $v_{cy}=[\dot{y}_{1}+\dot{y}_{2}]/2$. This figure corresponds to the interaction of two identical and stable single peak solitons with $B_3=0.7$ and $\delta=3.3$ [c.f. Fig. 2 in the main article].\label{fs2}}
\end{center}
\end{figure*}
The only time dependences in Eq. 21 are found in $\dot{x}_{1,2}\equiv\partial_tx_{1,2}$ and $\dot{y}_{1,2}\equiv\partial_ty_{1,2}$, which represent the soliton motions relative to $v_x$ and $v_y$, respectively. The right hand side of Eq. \ref{eq21} clearly states that the soliton motions $\dot{x}_{1,2},\dot{y}_{1,2}$ are driven by the correction $\vec{\chi}$ and the bare interaction between $S_1$ and $S_2$, contained in $\vec{\mathcal{K}}$. In deriving Eq.\ref{eq21}, the time derivatives $\partial_tS_1$, $\partial_tS_2$, and $\partial_t\chi$ have been neglected. Neglecting $\partial_tS_{1,2}$ is easily justified for stable solitons that are far enough from instability thresholds, so they exhibit no breathing nor any dynamical behavior. The latter also justifies why $\partial_t\chi$ is neglected. In addition, the correction $\chi$ moves in space together with $S_{1,2}$, and hence, strictly peaking, the terms $\dot{x}\partial_x\chi+\partial_y\dot{y}\chi$ exist. The reason why these extra velocity terms are not taken into account in the above derivation is that around $S_q$, $\dot{x}\partial_{x_q}\chi+\dot{y}\partial_{y_q}\chi\approx\dot{x_q}\partial_{x_q}\chi+\dot{y_q}\partial_{y_q}\chi\ll\dot{x_q}\partial_{x_q}S_q+\dot{y_q}\partial_{y_q}S_q$ because $|\chi|\ll|S_q|$, and therefore the term $\dot{x_q}\partial_{x_q}S_q+\dot{y_q}\partial_{y_q}S_q$, present in Eq. 21, is the dominating one.

In this perturbation theory smallness of $\chi$ is intrinsically linked to a small spatial overlap between $S_1$ and $S_2$. In this scenario, $\hat{\mathcal{L}}$, has two neutral eigenvectors (associated to two zero eigenvalues) around each soliton $S_q$, given by $\vec{\phi}^{(1)}_q=\partial_{x_q}\vec{S}_q$ and $\vec{\phi}^{(2)}_q=\partial_{y_q}\vec{S}_q$, which represent adiabatic translations of the solitons $S_{1,2}$. Therefore,
the perturbation $\chi$ is naturally regarded as the superposition of these neutral modes. Existence of modes $\vec{\phi}_q$ implies existence of neutral modes for the adjoint operator $\hat{\mathcal{L}}^\dagger$, $\vec{\eta}^{(1)}_q$ and $\vec{\eta}^{(2)}_q$, which we can only compute numerically. These neutral modes $\vec{\eta}^{(1)}_q$, $\vec{\eta}^{(2)}_q$ constitute the Kernel of $\hat{\mathcal{L}}^\dagger$ and thus provide the solvability condition for Eq. \ref{eq21}:
\begin{eqnarray}
&& \langle\vec{\eta}_p^{(j)}\vert\sum_{q=1}^2\left[\dot{x}_q\partial_{x_q}+\dot{y}_q\partial_{y_q}\right]\vec{S_q}\rangle=\langle\vec{\eta}_p^{(j)}\vert\vec{\mathcal{K}}\rangle,\label{eq25}\\ &&
\langle\vec{\eta}\vert \vec{f}\rangle\equiv\int_{0}^{L_x}\mathrm{d}x\int_{0}^{L_y}\mathrm{d}y\ \vec{\eta}(x,y)^*\vec{f}(x,y)\label{eq26},
\end{eqnarray}
where the identity $\langle\vec{\eta}_p^{(j)}\vert\hat{\mathcal{L}}\vec{\chi}\rangle\equiv\langle\vec{\chi}\vert\hat{\mathcal{L}}^{\dagger}\vec{\eta}_p^{(j)}\rangle^{\dagger}\equiv0$, for $p,j=1,2$, has been used. $L_{x,y}$ denote cavity lengths along $x$, $y$ directions. Equation \ref{eq25} can be cast in the form of the algebraic system [c.f., Eq. (3) in the main article]:
\begin{eqnarray}
&& 
\hat{\mathrm{A}}[\dot{x}_1, \dot{x}_{2}, \dot{y}_{1}, 
\dot{y}_{2}]^T=\bf{b},\\ &&
\hat{\mathrm{A}}\equiv\\ && \nonumber
\left[\begin{array}{cccc}
\langle\vec{\eta}_1^{(1)}\vert\partial_{x_1}\vec{S_1}\rangle &
\langle\vec{\eta}_1^{(1)}\vert\partial_{y_1}\vec{S_1}\rangle
&
\langle\vec{\eta}_1^{(1)}\vert\partial_{x_2}\vec{S_2}\rangle
&
\langle\vec{\eta}_1^{(1)}\vert\partial_{y_2}\vec{S_2}\rangle
\\
\langle\vec{\eta}_1^{(2)}\vert\partial_{x_1}\vec{S_1}\rangle &
\langle\vec{\eta}_1^{(2)}\vert\partial_{y_1}\vec{S_1}\rangle
&
\langle\vec{\eta}_1^{(2)}\vert\partial_{x_2}\vec{S_2}\rangle
&
\langle\vec{\eta}_1^{(2)}\vert\partial_{y_2}\vec{S_2}\rangle
\\
\langle\vec{\eta}_2^{(1)}\vert\partial_{x_1}\vec{S_1}\rangle &
\langle\vec{\eta}_2^{(1)}\vert\partial_{y_1}\vec{S_1}\rangle
&
\langle\vec{\eta}_2^{(1)}\vert\partial_{x_2}\vec{S_2}\rangle
&
\langle\vec{\eta}_2^{(1)}\vert\partial_{y_2}\vec{S_2}\rangle
\\
\langle\vec{\eta}_2^{(2)}\vert\partial_{x_1}\vec{S_1}\rangle &
\langle\vec{\eta}_2^{(2)}\vert\partial_{y_1}\vec{S_1}\rangle
&
\langle\vec{\eta}_2^{(2)}\vert\partial_{x_2}\vec{S_2}\rangle
&
\langle\vec{\eta}_2^{(2)}\vert\partial_{y_2}\vec{S_2}\rangle
\end{array}\right]
\\&&
{\bf{b}}\equiv[
\langle\vec{\eta}_1^{(1)}\vert\overrightarrow{\mathcal{K}}\rangle,
\langle\vec{\eta}_1^{(2)}\vert\overrightarrow{\mathcal{K}}\rangle,
\langle\vec{\eta}_2^{(1)}\vert\overrightarrow{\mathcal{K}}\rangle,
\langle\vec{\eta}_2^{(2)}\vert\overrightarrow{\mathcal{K}}\rangle
]^T.
\end{eqnarray}
The above system is solved with the Cramer's rule for each value of the soliton displacements $\Delta x$, $\Delta y$, and as a result we obtain the four scalar fields $\dot{x}_1(\Delta x,\Delta y)$, $\dot{y}_1(\Delta x,\Delta y)$, $\dot{x}_2(\Delta x,\Delta y)$, $\dot{y}_2(\Delta x,\Delta y)$.

Two-soliton clusters therefore exist for offsets $\{\Delta x,\Delta y\}$ such that the solitons $S_1$ and $S_2$ move with equal speeds, so $\Delta v_x\equiv\dot{x}_{1}-\dot{x}_{2}=0$ and $\Delta v_y\equiv\dot{y}_{1}-\dot{y}_{2}=0$. Figures \ref{fs2}(a,b) show $\Delta v_y$ and $\Delta v_x$, respectively, for the case of two interacting single peak and stable solitons with $B_3=0.7$ and $\delta=3.3$. The contour lines mark the zeros of these velocity shifts. These two contours are shown simultaneously in Fig. \ref{fs2}(c), and their intersections, i.e., two-soliton cluster solutions, are marked with various symbols, representing different stability features of the clusters, that can be only understood from Figs. \ref{fs2}(a,b). Stability against soliton-soliton forces is easily understood by looking at the vector field $\vec{F}=\textbf{e}_x{\Delta v_x}+\textbf{e}_y\Delta v_y$, where $\textbf{e}_x$, $\textbf{e}_y$ are unitary vectors along $\Delta x$, $\Delta y$, which is visualized by combining Figs. \ref{fs2}(a,b). For example, rhomboids in Fig. \ref{fs2}(c) correspond to unstable clusters because in their neighborhood $\vec{F}$ points away from their locations, as sketched on the right panel of Fig. \ref{fs2}(c). Similarly, hollow squares mark clusters which are stable only to small transverse displacements $\Delta y$ (and unstable against longitudinal displacements $\Delta x$), while hollow circles correspond to clusters which are stable only against small longitudinal displacements. Absolute stability of the clusters is only achieved at the off-axis locations marked by the solid circles. The trajectories in the plane $\Delta x,\Delta y$ that are tangent to $\vec{F}$ are streamlines describing the relative motion of the interacting solitons $S_1$ and $S_2$, as shown in Fig. 2(a) in the main article.

Cluster velocities are given by $v_{cx}=[\dot{x}_{1}+\dot{x}_{2}]/2$ and $v_{cy}=[\dot{y}_{1}+\dot{y}_{2}]/2$, shown in Figs. \ref{fs2}(e) and (d), respectively. 
Figures \ref{fs2}(d,e) also show the locations for two stable clusters (red dots). One remarkable feature of this theory is the ability to predict transverse drifts. Figure \ref{fs2}(d) predicts that the the cluster offset by $\Delta_x\approx-5.56,\Delta_y\approx-2.40$ has a transverse velocity $v_{cy}\approx3\times10^{-3}$, in very good agreement with the value $v_y=3.0085\times10^{-3}$ computed numerically with the modified Newton method explained in the preceding section (c.f., Fig. \ref{fs1} at $\delta=3.3$).
%
\begin{figure*}
\begin{center}
\includegraphics[width=.99\textwidth]{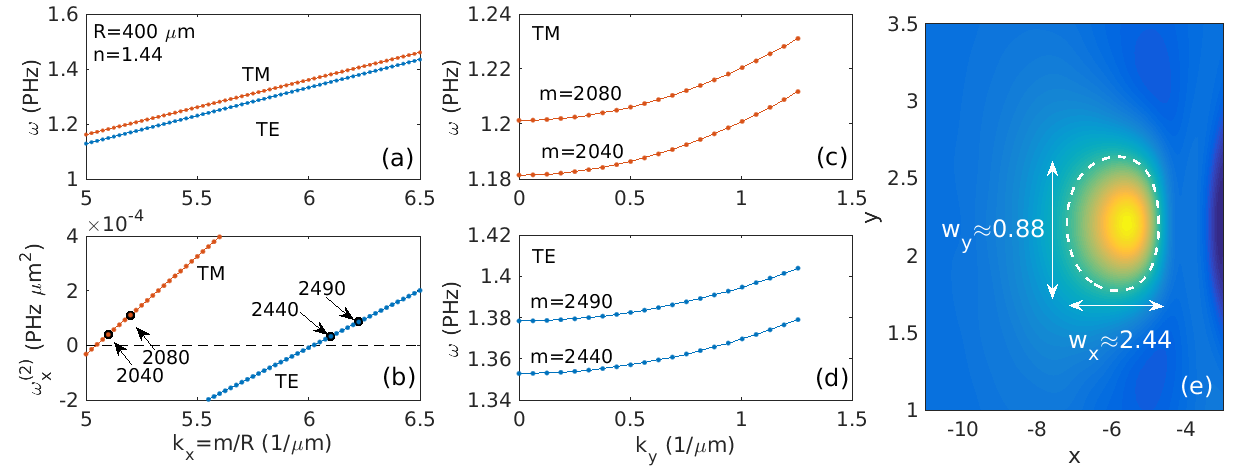}
\caption{Dispersion relations $\omega$ vs $k_x$ (a) and $\omega$ vs $k_y$ (c,d) of the fundamental TE and TM modes of a hollow cylinder with $R=400\ \mu$m, wall thickness of $1\ \mu$ and refractive index close to silica glass, $n=1.44$. (b) GVD coefficient $\omega^{(2)}_x\equiv\partial^2\omega/\partial k_x^2$ along the $x$ coordinate. (e) Zoom on soliton profile around its peak amplitude for $B_3=0.7$, $\delta=3.3$ showing soliton widths in $x$ and $y$ directions.\label{fs3}}
\end{center}
\end{figure*}
\section{ESTIMATES OF THE LONGITUDINAL AND TRANSVERSE SOLITON WIDTHS}
Soliton widths are calculated from the dispersion coefficients along $x$ and $y$. To this end we have used a Maxwell mode solver to compute typical dispersion profiles $\omega(k_x,k_y)$ for typical micro-cavity parameters. Figure \ref{fs3}(a) shows dispersion $\omega(k_x,0)$ for the fundamental TE and TM modes of a hollow cylindrical cavity [c.f Fig. 1(b) in the main article] of radius $R=400\ \mu$m, with a wall thickness of $1\ \mu$m, and refractive index $n=1.44$, close to silica for wavelengths of $1-2\ \mu$m. We consider the cavity to have a $Q-$factor around $10^6$, which corresponds to $\gamma\approx0.02$ in Eq. \ref{eq1}. Modes in Fig. \ref{fs3}(a) follow circular trajectories and have zero transverse motion. Group velocity is then computed as $v_g\equiv\omega^{(1)}_x\equiv\partial\omega(k_x,0)/\partial k_x\approx2\times10^8$ m/s corresponding to a roundtrip time $\tau=2\pi R/v_g\approx12$ ps for both modes in the plotted $k_x$ interval. Group velocity dispersion (GVD) along $x$ is computed as $\omega^{(2)}_x\equiv\partial^2\omega(k_x,0)/\partial k_x^2$ and shown in Fig. \ref{fs3}(b) for the two modes. The modal numbers $m=2040$, $2080$ on the TM branch, and $m=2440$, $2490$ on the TE branch, delimit the intervals where the dispersion satisfies the condition $\omega^{(3)}_x/(6\omega^{(2)}_x)\times\sqrt{\gamma/(\omega^{(2)}_x\tau})\in [0.16,0.7]$, which closely corresponds to the dispersion landscape assumed in the main article. The TM modes have a wavelength around $\lambda_0\approx1.56\ \mu$m while for the TE modes $\lambda_0\approx1.36\ \mu$m. Figures \ref{fs3}(c,d) show the dispersions of the above selected modes around their cut-off frequency, i.e., $\omega$ $vs$ wavenumber along the vertical direction, $k_y$, at fixed $m$. The GVD along the transverse direction is clearly anomalous $\omega^{(2)}_y\equiv\partial^2\omega(k_x,k_y)/\partial k_y^2>0$, as we described in the context of Eq. (1) in the main article.

Numerical data in Figs.\ref{fs3}(b-d) give the GVD along the longitudinal, $\omega_{x}^{(2)}\equiv\partial^2\omega/\partial k_{x}^2$ and transverse, $\omega_{y}^{(2)}\equiv\partial^2\omega/\partial k_{y}^2$, directions for the selected TE and TM modes:
$\omega^{(2)}_{x,TE}\in[0.033,0.086]$ $\mu$m$^2/$ps,
$\omega^{(2)}_{x,TM}\in[0.04,0.11]$ $\mu$m$^2/$ps,
$\omega^{(2)}_{y,TE}\approx32$ $\mu$m$^2/$ps,
$\omega^{(2)}_{y,TM}\approx37$ $\mu$m$^2/$ps,
$\omega^{(3)}_{x,TE}=0.42$ $\mu$m$^3/$ps,
$\omega^{(3)}_{x,TM}=0.72$ $\mu$m$^3/$ps. Therefore the $y$ and $x$ coordinates in Eq. \ref{eq1} are \textit{out of scale} by a factor $f\equiv\sqrt{{\omega^{(2)}_{y}}/\omega^{(2)}_{x}}$ which varies from $\approx18$ to $\approx30$ for the values $B_3=0.2$ and $B_3=0.7$ in the main article, respectively. Therefore, the coordinate $\tilde{y}\equiv fy$ is to scale with $x$. Far from the zero GVD ($B_3\approx0$) $f$ would also give the soliton width ratio along $x$ and $\tilde{y}$. Note $f=1$ for homogeneous media where dispersion (diffraction) is the same in all directions. Close to the zero GVD $B_3$ breaks isotropy between $x$ and $\tilde{y}$ so it is expected that $B_3$ will impact considerably the soliton widths. Hence estimates must be done from the numerically computed solutions.

Figure \ref{fs3}(e) shows a zoom around a soliton from Fig.3(b) in the main article (for $B_3=0.7$). The contour level placed at the half of the maximum soliton amplitude gives the widths $\mathrm{w}_x\approx2.44$ and $\mathrm{w}_y\approx0.88$, which yields $\mathrm{w}_{\tilde{y}}=f\mathrm{w}_y\approx26$. Therefore, physical transverse sizes of this soliton is $26/2.44\approx10.7$ times larger than longitudinal sizes (which differs substantially from the factor $f\approx30$ given above). In this example example, $w^{(2)}_x\approx0.04$ $\mu$m$^2$/ps giving $B_2\equiv\omega^{(2)}\tau/[(2\pi R)^2]\approx 7.6\times10^{-8}$. Hence the longitudinal soliton width  $\mathrm{w}_x\approx2.44$ corresponds to an angular width of $\Delta_\mathrm{\theta}=2\pi\mathrm{w}_x\sqrt{B_2/\gamma}\approx3\times10^{-2}$ rad, a duration of $\Delta_\tau=\tau\mathrm{w}_x\sqrt{B_2/\gamma}\sim60$ fs, a physical longitudinal length of $\Delta_x=\Delta\mathrm{\theta}R\approx12\ \mu$m, and a physical transverse size $\Delta_y=10.7\Delta_x\approx130\ \mu$m.

According to the above estimates, the largest of our clusters, shown in Fig. 5(e) in the main article, has a transverse size of $\approx20$ corresponding to a physical size of $20f\Delta_x/\mathrm{w}_x\approx3$ mm and hence this is the transverse length over which the external pump should remain approximately constant. Therefore, the pump geometry in a realistic experiment should consist on a planar waveguide placed in proximity to the cylinder's tangent.

\begin{figure}
\begin{center}
\includegraphics[width=.49\textwidth]{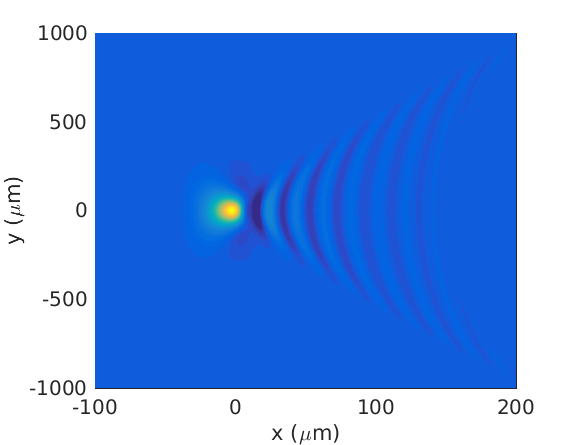}
\caption{Soliton profile in physical space corresponding to $B_3=0.16$, $\delta=2.8$ in the main article [c.f. Fig.1(a)].\label{fs4}}
\end{center}
\end{figure}

The transverse size of the CSCs can be tuned through the ratio $\Delta_y/\Delta_x$, which is sensitive to the cavity dispersion and other parameters, such as detuning, $\delta$. As an example, the above estimates applied to the soliton with $B_3=0.16$ and $\delta=2.8$ in the main article (and shown in Fig. \ref{fs4} in physical units)
, lead to $\Delta_\tau\approx106$ fs, $\Delta_x\approx22\ \mu$m, and $\Delta_y\approx100\ \mu$m$\approx4.5\Delta_x$. Optimal width ratios will depend on the particular scope and we did not attempt to address such issue in the present work.
%

\begin{thebibliography}{10}
\bibitem{mynote}
We scanned intensively the parameter space in $3D$, i.e., replacing $\partial_y^2\rightarrow\partial_y^2+\partial_z^2$ in Eq. \ref{eq1} and found that all radiating bullets with radial symmetry were clearly unstable.
\end{thebibliography}
\end{document}